\documentclass[aps,twocolumn,showpacs,amsmath,amssymb,pre]{revtex4}

\usepackage{graphicx}
\usepackage{dcolumn}
\usepackage{bm}
\usepackage{epsf}
\usepackage{amsmath}
\usepackage{amssymb}
\usepackage{color}

\newcommand{\bra}[1]{\langle #1|}
\newcommand{\ket}[1]{|#1\rangle}

\newcommand{\la}{\left\langle}
\newcommand{\ra}{\right\rangle}

\newcommand{\bla}{bla\\bla\\bla\\bla\\bla}

\newcommand{\PRA}{Phys. Rev. A }

\newcommand{\PRE}{Phys. Rev. E }
\newcommand{\PRX}{Phys. Rev. X }
\newcommand{\PRL}{Phys. Rev. Lett. }
\newcommand{\EPL}{EPL }

\newcommand{\NJP}{New. J. Phys. }
\newcommand{\NC}{Nat. Comm. }

\begin{document}

\title{Performance of superadiabatic quantum machines}
\author{Obinna Abah}
\affiliation{Department of Physics, Friedrich-Alexander-Universit\"at Erlangen-N\"urnberg, D-91058 Erlangen, Germany}
\author{Eric Lutz}
\affiliation{Department of Physics, Friedrich-Alexander-Universit\"at Erlangen-N\"urnberg, D-91058 Erlangen, Germany}


\begin{abstract}
We investigate the performance of a quantum thermal machine operating in finite time based on shortcut-to-adiabaticity techniques. We  compute efficiency and power for a quantum harmonic Otto engine by taking the energetic cost of the superadiabatic driving explicitly into account. We further derive generic upper bounds on both quantities, valid for any heat engine cycle, using the notion of quantum speed limits for driven systems. We demonstrate that these quantum bounds are tighter than those stemming from the second law of thermodynamics.

\end{abstract}

\maketitle

Superadiabatic (SA) techniques  allow the engineering of  adiabatic dynamics in  finite time. While truly adiabatic transformations require infinitely slow driving, transitionless protocols may be implemented at finite speed by adding properly designed time-dependent terms $H_\text{SA}(t)$ to the Hamiltonian of a system  \cite{dem03,ber08}. By suppressing nonadiabatic excitations, these fast processes reproduce the same final state    as that of adiabatic driving. In that sense, they provide a shortcut to adiabaticity. In the last few years, there has been remarkable progress, both theoretical \cite{dem03,ber08,mug10,che10,iba12,cam13,def14} and experimental \cite{cou08,sch10,sch11,bas12,bow12,wal12,zha13,du16,an16,mar16}, in developing superadiabatic methods for  quantum and classical systems (see  Ref.~\cite{tor13} for a review). Successful applications include  high-fidelity driving of a BEC \cite{bas12}, fast transport of trapped ions \cite{bow12,wal12}, fast  adiabatic passage using a single spin in diamond \cite{zha13} and cold atoms \cite{du16}, as well as swift equilibration of a Brownian particle \cite{mar16}. 

Superadiabatic protocols have recently been   extended   to thermal machines as a means  to enhance their performance. Classical \cite{den13,tu14} and  quantum \cite{cam14,bea16} single particle heat engines, as well as multiparticle quantum motors \cite{bea16,jar16,cho16} have been theoretically investigated.
 Nonadiabatic transitions are  well-known sources  of entropy production that reduce  the efficiency of thermal machines \cite{and84,and11,fel00}. Successfully suppressing them using superadiabatic methods thus appears a promising  strategy to boost their work and power output.
 
  However, a crucial point that needs to be addressed in order to assess the usefulness of shortcut techniques in thermodynamics is the proper computation of the efficiency of a superadiabatic engine. Since the excitation suppressing term $H_\text{SA}(t)$ in the Hamiltonian is often assumed to be zero at the begin and at  the end of a transformation \cite{tor13}, its  work contribution vanishes. The energetic cost of the additional superadiabatic driving is therefore commonly not included in the calculation of the efficiency \cite{den13,tu14,cam14,bea16,jar16,cho16}.  As a result,  the latter quantity reduces to the adiabatic efficiency, even for fast nonadiabatic driving of the machine: the superadiabatic driving thus  appears to be for free. This situation is somewhat reminiscent of the power of a periodic signal which is zero at the beginning and at the end of one period. While the  instantaneous power vanishes at the end of the interval, the actual power of the signal is  given  by the non-zero time-averaged power \cite{hal13}. As a matter of  fact,  the energetic cost of superadiabatic protocols was lately defined in universal quantum computation and adiabatic gate teleportation models as the time-averaged norm of  the superadiabatic Hamiltonian $H_\text{SA}(t)$ \cite{san15,san16} (see also Refs.~\cite{zhe15,cou16,cam16,fun16}). However, the chosen Hilbert-Schmidt norm is related to the variance of the energy and not to its mean \cite{def13}. It is hence of limited relevance to the investigation of the energetics of a heat engine. 
 
 In this paper, we evaluate the performance of a superadiabatic thermal machine by properly taking  the energetic cost of the transitionless driving into account. We consider commonly employed local counterdiabatic (LCD) control techniques  \cite{iba12,cam13,def14} which have latterly been successfully implemented experimentally in Refs.~\cite{sch10,sch11,bas12}. We evaluate both efficiency and power for a paradigmatic harmonic quantum Otto engine \cite{kos84,gev92,lin03,rez06,qua07,aba12}. We explicitly compute the cost of the superadiabatic protocol as the time-averaged expectation value of the Hamiltonian $H_\text{SA}(t)$ for compression and expansion phases of the engine cycle. We find that the energetic cost of the superadiabatic driving exceeds the potential work gain for moderately rapid protocols. Superadiabatic engines may therefore only outperform traditional quantum motors for very fast cycles, albeit with an efficiency much smaller than the corresponding adiabatic efficiency.  We additionally derive generic upper bounds on both superadiabatic efficiency and power, valid for any heat engine cycle, based on the concept of quantum speed limit times for driven unitary dynamics \cite{def13a}. We demonstrate that these quantum bounds are tighter than conventional bounds that follow from the second law of thermodynamics.

\begin{figure}[t]
\begin{center}
\includegraphics[width=1 \linewidth]{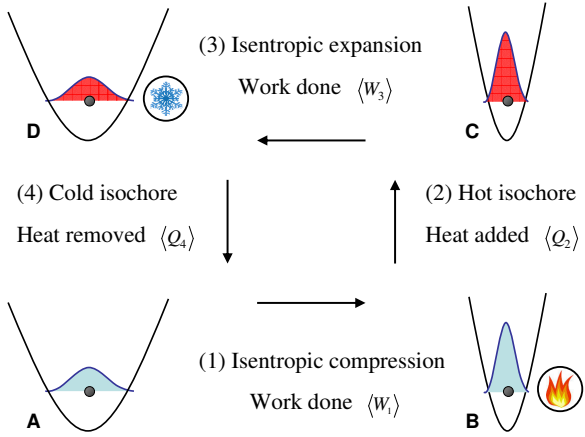}
\end{center}
\caption{Quantum Otto engine for a harmonic trap with time-dependent frequency. The cycle consists of four consecutive steps: (1) isentropic compression, (2) isochoric heating, (3) isentropic expansion and (4) isochoric cooling. Work is produced during the first and third unitary strokes, while heat is absorbed from the hot reservoir during the heating phase (2).  }
\label{fig1}
\end{figure}

\textit{Quantum Otto engine.}
 We consider a quantum engine whose working medium is a  harmonic oscillator with time-dependent frequency $\omega_t$. The corresponding Hamiltonian is of the usual form, $H_0(t) = p^2/(2 m) + m\omega_t^2 x^2/2$, where $x$ and $p$ are the position and momentum operators of an oscillator of mass $m$.  The Otto cycle consists of four consecutive steps as shown in Fig.~\ref{fig1} \cite{kos84,gev92,lin03,rez06,qua07,aba12}:
(1) \textit{Isentropic compression} $A\rightarrow B$: the  frequency is varied from $\omega_1$ to $\omega_2$ during time $\tau_1$ while the system is isolated. The evolution is unitary and the von Neumann entropy  is  constant. (2) \textit{Hot isochore} $B\rightarrow C$: the oscillator is weakly coupled to a bath at inverse temperature $\beta_2$ at fixed frequency and  thermalizes to state C during time $\tau_2$.
(3) \textit{Isentropic expansion} $C\rightarrow D$: the frequency is changed back to its initial value during time $\tau_3$ at constant von Neumann entropy.
(4)  \textit{Cold isochore} $D\rightarrow A$:  the system is weakly coupled to a bath at inverse temperature $\beta_1>\beta_2$  and  relaxes to  state A during $\tau_4$ at fixed frequency. We will assume, as commonly done \cite{kos84,gev92,lin03,rez06,qua07,aba12}, that the thermalization times $\tau_{2,4}$ are much shorter than the compression/expansion times $\tau_{1,3}$. The total cycle time is then  $\tau_\text{cycle}= \tau_1+ \tau_3=2 \tau$ for equal step duration.

In order to evaluate the performance of the Otto engine, we need to compute work and heat for each of the above steps. Work is performed during the first and third unitary strokes, whereas heat is exchanged with the baths during the isochoric thermalization phases two and four. The mean work may be calculated by using the exact solution of the Schr\"odinger equation for the parametric oscillator for any given frequency modulation \cite{def08,def10}. For the compression/expansion steps, it is given by \cite{aba12},
\begin{eqnarray}
\la W_1\ra &=& \frac{\hbar}{2} (\omega_2Q^\ast_1 - \omega_1) \coth\left(\frac{\beta_1\hbar\omega_1}{2}\right),\\
\la W_3\ra &=& \frac{\hbar}{2} (\omega_1 Q^\ast_3 - \omega_2) \coth\left(\frac{\beta_2\hbar\omega_2}{2}\right),
\end{eqnarray}
where we have introduced the dimensionless adiabaticity parameter $Q^*_{i}$ $(i=1,3)$ \cite{hus53}. It is defined as the ratio of the mean energy and the corresponding adiabatic mean energy and is thus equal to one for adiabatic processes \cite{def10}. Its explicit expression for any frequency modulation $\omega_t$ may be found in Refs.~\cite{def08,def10}.  Furthermore, the mean heat absorbed from the hot bath  reads \cite{aba12},
 \begin{equation}
\la Q_2\ra = \frac{\hbar \omega_2}{2} \left[\coth\left(\frac{\beta_2\hbar\omega_2}{2}\right) - Q^\ast_1 \coth\left(\frac{\beta_1 \hbar \omega_1}{2}\right) \right]. \label{3}
\end{equation}
For an engine,  the produced work  is negative, $\la W_1\ra +\la W_3\ra <0$, and the absorbed heat is positive, $\la Q_2\ra >0$.

\begin{figure}[t]
\begin{center}
\includegraphics[width=.96\linewidth]{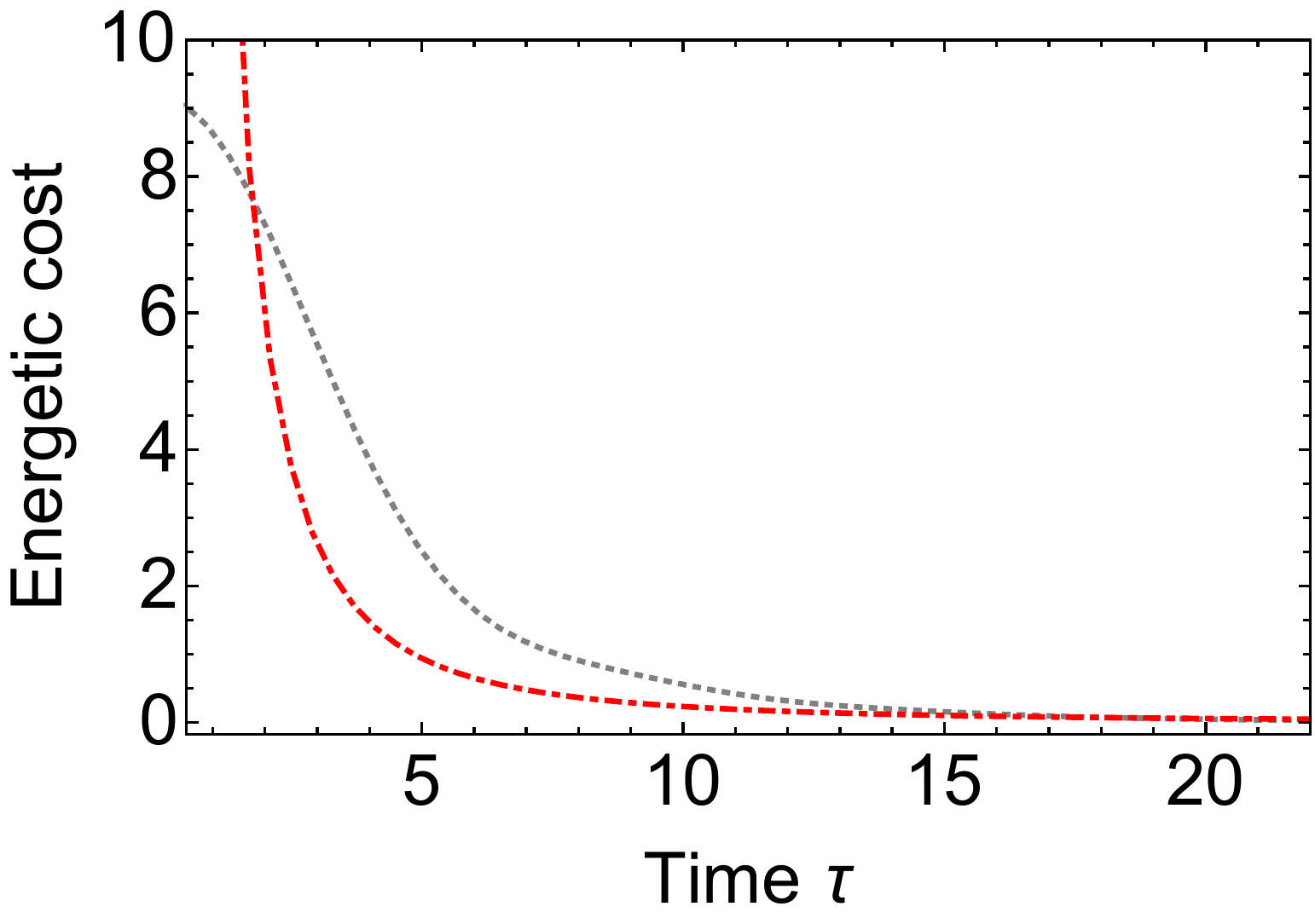}
\end{center}
\caption{  Energetic cost of the superadiabatic driving  $\la H^1_\text{SA}\ra_\tau+ \la H^3_\text{SA}\ra_\tau$, each defined as the time average  of Eq.~(8),  for  the compression and expansion steps (1) and (3) (red dotted-dashed) as a function of the  driving time $\tau$.  The corresponding nonadiabatic work $\la W_1 \ra_\text{NA}+ \la W_3 \ra_\text{NA}$, defined as the difference between the actual  and the adiabatic work, is shown for comparison (grey dotted). Parameters are $\omega_1 = 0.32$, $\omega_2 = 1$, $\beta_1 =  0.5$ and $\beta_2 = 0.05$.}
\label{fig2}
\end{figure}

\textit{Superadiabatic driving.} The  compression and expansion  phases (1) and (3) may be sped up, while suppressing unwanted nonadiabatic transitions, by adding a local harmonic potential $H_\text{SA}$  to the system Hamiltonian $H_0$. The local counterdiabatic Hamiltonian may then written in  the form $H_\text{LCD}(t) = H_0(t) +  H_\text{SA}(t) $ with \cite{iba12,cam13,def14},
 \begin{equation}
  H_\mathrm{SA} =   \frac{m}{2} \left ( \Omega^2_t -\omega^2_t \right ) x^2= \frac{m}{2} \left(- \frac{3 \dot{\omega}_t^2}{4 \omega_t^2} + \frac{\ddot{\omega}_t}{2 \omega_t}\right) x^2. \label{4}
\end{equation} 
Boundary conditions ensuring that $H_\text{SA}(0,\tau)=0$ at the beginning and at the end of the driving are given by,   
 \begin{equation}
\begin{array}{lcr}
\omega(0) = \omega_i,& \dot{\omega}(0) = 0 ,& \ddot{\omega}(0) = 0,\\
\omega(\tau) =\omega_f,& \dot{\omega}(\tau) = 0 ,& \ddot{\omega}(\tau) = 0, \label{5}
\end{array}
\end{equation}
where $\omega_{i,f}= \omega_{1,2}$ denote the respective initial  and final frequencies of  the compression/expansion steps. The  conditions \eqref{5} are, for instance,  satisfied by \cite{iba12,cam13,def14},
\begin{equation}
\omega_t= \omega_i + 10(\omega_f - \omega_i)s^3 - 15(\omega_f-\omega_i)s^4+ 6(\omega_f - \omega_i) s^5,
\end{equation}  
with $s = t/\tau$. Note that $\Omega_t^2 > 0$ to avoid trap inversion. 
Implementing the superadiabatic driving \eqref{4} leads to a unit adiabaticity parameter, $Q^*_i(\tau)=1$ $(i=1,3)$. As a consequence, the work performed in finite time during the two compression/expansion phases is equal to the adiabatic work, $  \la W_1\ra_\text{SA} =\la W_1\ra_\text{AD}$ and  $  \la W_3\ra_\text{SA} =\la W_3\ra_\text{AD}$.

\textit{Efficiency of the superadiabatic engine.} We define the efficiency of the superadiabatic motor as,
\begin{equation}
\eta_\text{SA} = \frac{\mathrm{energy \,output}}{\mathrm{energy\, input}} = \frac{-(\la W_1\ra_\text{SA}+\la W_3\ra_\text{SA})}{\la Q_2\ra + \la H^1_\mathrm{SA}\ra_\tau + \la H^3_\mathrm{SA}\ra_\tau}. \label{7}
\end{equation} 
In the above expression,   the energetic cost of the transitionless driving is taken into account by including the time-average, $\la H^i_\text{SA}\ra_\tau = (1/\tau) \int_0^\tau dt \la H^i_\text{SA}(t)\ra$  $(i=1,3)$, of the local potential \eqref{4} for the compression/expansion steps.  Equation \eqref{7} reduces to the adiabatic efficiency $\eta_\text{AD}$ in the absence of these two contributions.  For further reference, we also introduce the usual nonadiabatic efficiency of the engine, $\eta_\text{NA} = -(\la W_1\ra+\la W_3\ra)/\la Q_2\ra$,  based on the formulas (1)-(3) without any shortcut.

The expectation value of the local counterdiabatic potential \eqref{4} may be calculated explicitly for an initial thermal state in terms of the initial energy of the system $\la H_0(0)\ra$. We find (see Supplemental Material \cite{sup}), 
\begin{equation}
\la H_\mathrm{SA}(t) \ra = \frac{\omega_t}{\omega_i} \la H_0(0) \ra\left[ -\frac{\dot{\omega}_t^2}{4 \omega_t^4} + \frac{\ddot{\omega}_t}{4 \omega_t^3}\right]. \label{8}
\end{equation}
We  use Eq.~\eqref{8}  to numerically compute the time averages $\la H^i_\text{SA}\ra_\tau$ $(i=1,3)$ for compression/expansion that are needed to evaluate the superadiabatic efficiency \eqref{7}.

\begin{figure}[t]
\begin{center}
\includegraphics[width=1\linewidth]{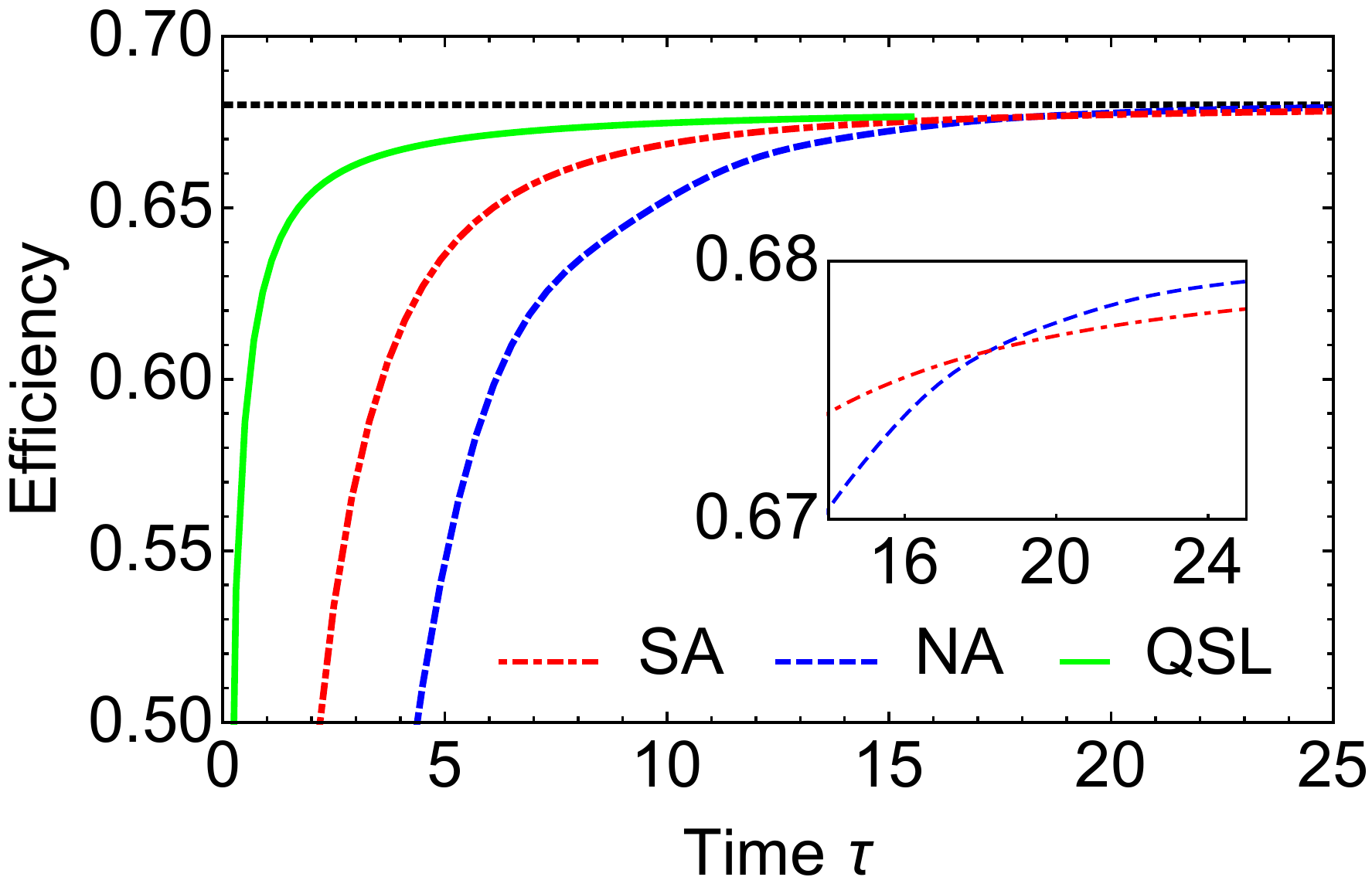}
\end{center}
\caption{Superadiabatic efficiency $\eta_\text{SA}$ (red dotted-dashed), Eq.~(7), together with the nonadiabatic efficiency $\eta_\text{NA}$ (blue dashed) and the adiabatic efficiency $\eta_\text{AD}$ (black dotted)  as a function of the  time $\tau$. The green solid line shows the quantum speed limit bound (11). Same parameters as in  Fig.~2.}
\label{fig3}
\end{figure}

 Figure 2 shows, as an illustration, the energetic cost of the superadiabatic driving  $\la H^1_\text{SA}\ra_\tau+ \la H^3_\text{SA}\ra_\tau$, for the compression and  expansion  steps (1) and (3) as a function of the driving time $\tau$. We also display,  for comparison, the corresponding nonadiabatic work $\la W_1 \ra_\text{NA}+ \la W_3 \ra_\text{NA}$, defined as the difference between the actual work and the adiabatic work, $\la W_i \ra_\text{NA}= \la W_i \ra - \la W_i\ra_\text{AD}$ $(i=1,3)$; this quantity  measures the importance of nonadiabatic excitations induced by fast protocols and is often referred to as  internal friction \cite{fel00,rez06,cam14,zhe15}. We observe that the time-averaged superadiabatic energy (red dotted-dashed) increases significantly with decreasing process time as expected. This increase is much faster than that of the nonadiabatic work  $\la W_1 \ra_\text{NA}+\la W_3 \ra_\text{NA}$ (grey dotted). Eventually, for very rapid driving, the energetic price of the shortcut will dominate nonadiabatic energy losses.

Figure 3 exhibits the superadiabatic efficiency $\eta_\text{SA}$ (red dotted-dashed), Eq.~\eqref{7}, as a function of the  driving time $\tau$, together with the adiabatic efficiency $\eta_\text{AD}$ (black dotted) and the nonadiabatic efficiency $\eta_\text{NA}$ (blue dashed). Three points are worth emphasizing: i) if the energetic cost of the shortcut is not included, the superadiabatic efficiency is equal to the maximum possible value given by the constant adiabatic efficiency $\eta_\text{AD}$, as noted in Refs.~{\cite{den13,tu14,cam14,bea16,jar16,cho16}, ii) by contrast, if that energetic cost is properly taken into account,  the superadiabatic efficiency $\eta_\text{SA}$ drops for decreasing $\tau$, reflecting the  sharp augmentation of the time-averaged superadiabatic energy seen in Fig.~2, iii)   we further observe that $\eta_\text{SA}<\eta_\text{NA}$ for large   time $\tau$, while $\eta_\text{SA}>\eta_\text{NA}$ only for small enough $\tau$ (inset). We can thus conclude that the superadiabatic driving is only of advantage for sufficiently short cycle durations. For large cycle times, the energetic cost of the shortcut  outweighs the work gained by emulating adiabaticity. 

Another benefit of the superadiabatic driving appears for very small time $\tau$. An examination of Eq.~\eqref{3} reveals that  the heat $\la Q_2\ra $ becomes negative for very strongly nonadiabatic processes, when $Q^*_1(\tau) > \coth({\beta_2\hbar\omega_2}/{2}) / \coth({\beta_1 \hbar \omega_1}/{2})$. In this regime, heat is pumped into the hot reservoir, instead of being absorbed from it, and the machine stops working as an engine \cite{obi16}. Since $Q^*_1(\tau) =1$ for all $\tau$ for the local counterdiabatic driving, this problem never occurs for the superadiabatic motor, even in the limit of very short cycles $\tau \rightarrow 0$.

\textit{Power of the superadiabatic engine.}
The power of the superadiabatic machine is given by,
\begin{equation}
P_\text{SA} = -\frac{\la W_{1}\ra_\text{SA} + \la W_{3}\ra_\text{SA}}{\tau_\text{cycle}}. \label{9}
\end{equation}
Since the superadiabatic protocol ensures adiabatic work output, $\la W_{i}\ra_\text{SA} = \la W_{i}\ra_\text{AD}$ $(i=1,3)$, in a shorter cycle duration $\tau_\text{cycle}$, the superadiabatic power $P_\text{SA}$ is always greater than the nonadiabatic power $P_\text{NA}=-(\la W_{1}\ra + \la W_{3}\ra)/{\tau_\text{cycle}}$ (see Fig.~4). This ability to considerably enhance the power of a thermal machine is  one of the true advantages of  the shortcut to adiabaticity approach. However, in view of the  discussion above, it is not possible to reach arbitrarily large power at maximum efficiency, as sometimes claimed \cite{den13,tu14,cam14,bea16}. This observation is in complete agreement with recent general proofs that forbid the simultaneous attainability of maximum power and maximum efficiency \cite{shi16}.

\begin{figure}[t]
\begin{center}
\includegraphics[width=.99\linewidth]{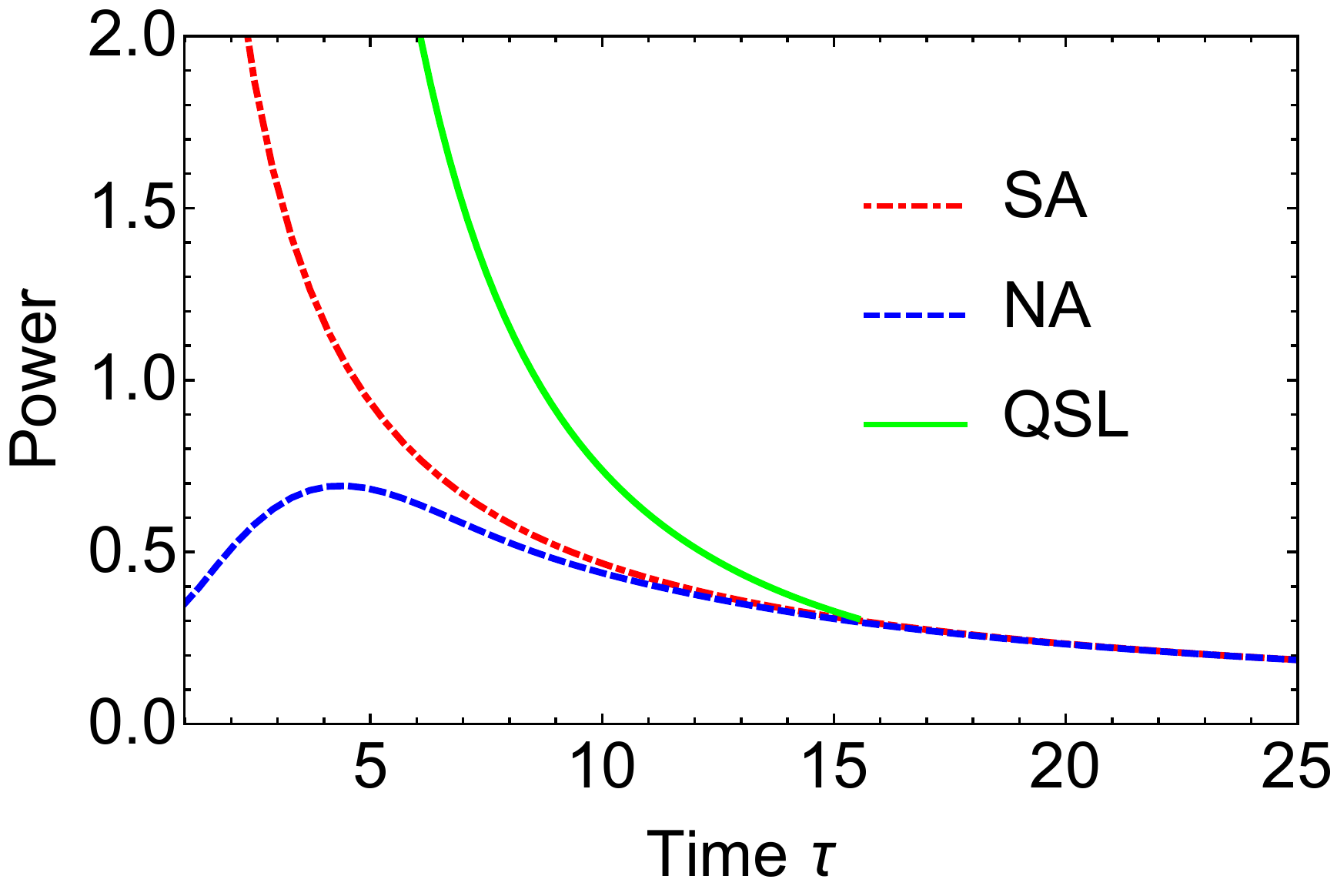}
\end{center}
\caption{Superadiabatic power $P_\text{SA}$ (red dotted-dashed), Eq.~(8), together with the nonadiabatic power  $P_\text{NA}$ (blue dashed)   as a function of the  driving time $\tau$. The green solid line shows the quantum speed limit bound (12). Same parameters as in  Fig.~2.}
\label{fig4}
\end{figure}

\textit{Universal quantum speed limit bounds.}
We finally derive generic upper bounds for both the superadiabatic efficiency \eqref{7} and the superadiabatic power \eqref{9}, based on the concept of quantum speed limits (see  Refs. \cite{ana90,vai91,uff93} and references therein). Contrary to classical physics, quantum theory limits the speed of evolution of a system  between given initial and final states. In particular, there exists a lower bound, called the quantum speed limit time $\tau_\text{QSL}\leq \tau$, on the time a system needs to evolve between these two states. An important restriction of the superadiabatic technique is the time required to successfully implement the counterdiabatic driving \eqref{4}, which depends on the first two time derivatives of the frequency $\omega_t$ \cite{com}. For this unitary driven dynamics, a Margolus-Levitin-type  bound on the evolution time reads \cite{def13a},
\begin{equation}
\tau \geq \tau_\text{QSL}= \frac{\hbar \mathcal{L}(\rho_i,\rho_f)}{\la H_\text{SA}\ra_\tau},  \label{10}
\end{equation}
where $\mathcal{L}(\rho_i,\rho_f)$ denotes the Bures angle between the initial and final density operators of the system \cite{def13b,sup} and 
$\la H_\text{SA}\ra_\tau$ the time-averaged superadiabatic energy \eqref{8}. We expect  Eq.~\eqref{10} to be  a proper bound for the compression/expansion phases, when the engine dynamics is dominated by the superadiabatic driving for small $\tau$.

To derive an upper bound on the superdiabatic efficiency \eqref{7}, we use  inequality \eqref{10} to obtain,
\begin{equation}
\eta_\text{SA} \leq \eta_\text{SA}^\text{QSL} = -\frac{\la W_1\ra_\text{AD}+\la W_3\ra_\text{AD}}{\la Q_2\ra +\hbar(\mathcal{L}_1+\mathcal{L}_3 )/\tau},\label{11}
\end{equation}
where $\mathcal{L}_i$ $(i=1,3)$ are the respective Bures angles for the compression/expansion steps. On the other hand, an upper bound on the superadiabatic power \eqref{9} is,
\begin{equation}
P_\text{SA}\leq P_\text{SA}^\text{QSL} = -\frac{\la W_{1}\ra_\text{AD} + \la W_{3}\ra_\text{AD}}{\tau^1_\text{QSL}+\tau^3_\text{QSL}}, \label{12}
\end{equation}
where $\tau^i_\text{QSL}$ $(i=1,3)$ are the respective  speed-limit bounds \eqref{10} for the compression/expansion phases.

The two quantum speed limit bounds \eqref{11} and \eqref{12} are shown in Figs.~(3)-(4) (green solid). We first  notice that the quantum bound \eqref{11} on the efficiency is sharper than the thermodynamic bound given by the constant adiabatic efficiency $\eta_\text{AD}$. Remarkably, quantum theory further imposes an upper bound  on  the power, whereas the second law of thermodynamics does not \cite{shi16}. Quantum thermodynamics hence establishes tighter bounds than classical thermodynamics.  The latter result may be understood by noting that thermodynamics does not have the notion of time scale, contrary to quantum mechanics. Finally, we stress that the two  speed limit bounds directly follow from  the definitions  of efficiency and power. As a result, they are independent of the thermodynamic cycle considered and generically apply to any quantum heat engine, not just to the quantum Otto motor.

\textit{Conclusions.}
We have performed a detailed study of both efficiency and power of a superadiabatic quantum heat engine. We have explicitly accounted for the energetic cost of the superadiabatic driving, defined as the time average of the local counterdiabatic potential. We have found that the efficiency of the engine markedly drops with decreasing cycle time. However, this drop is much slower than that of the nonadiabatic efficiency without the shortcut. As result, superadiabatic machines outperforms their conventional counterparts for very short cycles, when the work gain generated by the counterdiabatic driving outweighs its  energetic cost. We have additionally derived generic upper bound on superadiabatic efficiency and power based on the idea of quantum speed limits. These quantum bounds, valid for general thermal motors, are tighter than the usual bounds based on the second law of thermodynamics. We therefore expect them to be useful for future investigations of  thermal machines in the quantum regime.

\textit{Acknowledgments} This work was partially supported
by the EU Collaborative Project TherMiQ (Grant Agreement
618074) and the COST Action MP1209.

\section*{Supplemental Material}

\appendix{

\section{Local counterdiabatic energy}
We here present a derivation of the mean energy of the local counterdiabatic Hamiltonian $H_\text{LCD}$ and of the corresponding adiabaticity parameter $Q^*_\text{LCD}$ used  during the compression/expansion protocols. We consider a time-dependent harmonic oscillator with Hamiltonian, 
\begin{equation}
H_0(t) = \frac{p^2}{2 m} + \frac{m\omega_t^2 x^2}{2},
\label{s1}
\end{equation}
where $\omega_t$ is the time-dependent angular frequency, $m$  the mass, and ($p$, $x$)  the respective momentum and position operators. The initial energy eigenstates at $t=0$ with $\omega(0) = \omega_0$  in  coordinate representation are given by
\begin{equation}
\psi_n(x,0) = \frac{1}{\sqrt{2^n n!}} \left(\frac{m\omega_0}{\pi \hbar}\right)^{1/4} \exp\left(-\frac{m\omega_0}{2\hbar}x^2\right) \mathcal{H}_n\left(\sqrt{\frac{m\omega_0}{\hbar}}x\right),
\label{s2}
\end{equation}
 where $\mathcal{H}_n$ are  Hermite polynomials and  $E_n^0 = \hbar \omega_0 (n + 1/2)$ the corresponding energy eigenvalues. The instantaneous eigenstates and their corresponding eigenvalues are obtained by replacing $\omega_0$ with $\omega_t$.

The shortcut to adiabaticity may be implemented by adding a time-dependent counterdiabatic (CD) term to the system Hamiltonian (\ref{s1}) \cite{mug10,che10,iba12,cam13}:
\begin{equation}
H_\mathrm{SA}^\mathrm{CD}(t) = -\frac{\dot{\omega}_t}{4\, \omega_t}({x}{p}+ {p} {x}) = i\hbar \frac{\dot{\omega}_t}{4\omega_t} ({a}^2 -{a}^{\dagger 2}).
\label{s3}
\end{equation}
The last equality is obtained by expressing ${x} = \sqrt{\hbar/2m\omega_t}\,  ({a}^{\dagger} + {a})$ and ${p} = i\sqrt{\hbar m \omega_t/2}\,  ({a}^{\dagger} -{a})$, in terms of the annihilation and creation operators ${a}$ and ${a}^\dagger$. The total Hamiltonian $H_\mathrm{CD}(t) = H_\mathrm{0}(t) + H_\mathrm{SA}^\mathrm{CD}(t)$ is still quadratic in ${x}$ and ${p}$ and  may thus be considered that of a generalized harmonic oscillator \cite{ber85,che10}.  However, since the Hamiltonian (\ref{s3}) is a nonlocal operator, it is often convenient 
 to look for a unitarily equivalent Hamiltonian with a local   potential \cite{iba12,cam13}. Applying the canonical transformation, $U_x = \exp\left({i m \dot{\omega} x^2}/{4 \hbar \omega}\right)$, which cancels the cross terms $xp$ and $px$, to the Hamiltonian (\ref{s3}) leads to a new local counterdiabatic (LCD) Hamiltonian of the form \cite{iba12,cam13},
\begin{eqnarray}
H_\mathrm{LCD}(t) &=& U_x^\dagger (H_\mathrm{CD}(t) - i\hbar \dot{U}_x U_x^\dagger) U_x \nonumber \\ &=& \frac{p^2}{2 m} + \frac{m\Omega_t^2 x^2}{2},
\label{s4}
\end{eqnarray}
with the modified time-dependent (squared) frequency,
\begin{equation}
\Omega^2(t) = \omega_t^2 -\frac{3\dot{\omega}_t^2}{4\omega_t^2}+\frac{\ddot{\omega}_t}{2\omega_t}. 
\label{s5}
\end{equation}
This resulting  Hamiltonian  is local and still drives the evolution along the adiabatic trajectory of the system of interest. By demanding that $H_\mathrm{LCD} = H_\mathrm{0}$ at $t = {0,\tau}$ and imposing $\dot{\omega} (\tau) = \ddot{\omega} (\tau) = 0$, the final state is  equal for both dynamics, even in phase, and the final vibrational state populations coincide with those of a slow adiabatic process \cite{iba12}. It can be readily  shown that $\Omega^2(t)$ approaches $\omega^2(t)$ for very slow expansion/compression process \cite{cam13}.

Exact solutions of the  Schr\"odinger equation for a time-dependent harmonic oscillator have been extensively investigated \cite{lew69,ber84,loh09}. Following Lohe \cite{loh09}, a  solution  based on the invariants of motion is of the form, 
\begin{equation}
I(t) = \frac{b^2}{2 m} p^2 + \frac{m\dot{b}^2}{2}x^2 - \frac{b\, \dot{b}}{2}(p x + x p) + \frac{m \omega_0^2}{2b^2} x^2,
\label{s6}
\end{equation}
where $\omega_0$ is an arbitrary constant---a convenient  choice is to set $\omega_0 = \omega(0),  (\omega_0^2 > 0) $. The scaling factor $b = b(t)$ is a solution of the Ermakov differential equation,
\begin{equation}
\ddot{b} + \omega_t^2 b = \omega_0^2/b^3.
\label{s7}
\end{equation}
In the adiabatic limit, $\ddot{b} \simeq 0$ and 
\begin{equation}
b(t) \rightarrow b_{ad} = \sqrt{\frac{\omega_0}{\omega_t}}.
\label{s8}
\end{equation}
Equation (\ref{s7}) is valid for any given $\omega_t$ and its general solution can be constructed from the solutions $f(t)$ of the linear equation of motion for the classical time-dependent harmonic oscillator \cite{pin50},
\begin{equation}
\ddot{f} + \omega_t^2 f = 0,
\label{s9}
\end{equation}
according to $b^2/\omega_0 = f_1^2 + W^{-2}f_2^2$, where $f_1$, $f_2$ are independent solutions of Eq. (\ref{s9}) and the Wronskian $W[f_1,f_2] = f_1\dot{f}_2 - \dot{f}_1 f_2$ is a nonzero constant. We note that the Wronskian properties of Eq.~(\ref{s9}) can be used to show the equivalence of the adiabaticity parameter derived here and that of Husimi \cite{hus53,def10} (see Ref.~\cite{jar16}). The general solution of the time-dependent Schr\"odinger equation for the Hamiltonian (\ref{s4}) is hence,
\begin{widetext}
\begin{equation}
\Psi_n(x,t) = \frac{1}{\sqrt{2^n n!}} \left(\frac{m \omega_0}{\pi \hbar b^2}\right)^{1/4} \exp\left[\frac{i m \dot{b}}{2 \hbar b}x^2 - i\int_0^t \frac{\omega_0(n+1/2)}{b(t^\prime)^2}dt^\prime\right]  \exp\left(-\frac{m \omega_0}{2 \hbar b^2}x^2\right) \mathcal{H}_n\left(\sqrt{\frac{m\omega_0}{\hbar}}\frac{x}{b}\right).
\label{s10}
\end{equation}
\end{widetext}
The time-dependent energy eigenstates, $H_\text{LCD}|\Psi_n(x,t) \rangle = E |\Psi_n(x,t) \rangle $, are explicitly given by, 
\begin{eqnarray}
E &=& \bra{\Psi_n(x,t)} H_\mathrm{LCD} \ket{\Psi_n(x,t)} \nonumber \\ &=& \frac{E_n^0}{b^2} - \frac{m}{2}\left(b\ddot{b} - \dot{b}^2\right)  \bra{\Psi_n(x,0)} x^2 \ket{\Psi_n(x,0)} \nonumber \\ &+ &  \frac{\dot{b}}{2 b} \bra{\Psi_n(x,0)} x p + px \ket{\Psi_n(x,0)}.
\label{s11}
\end{eqnarray}

We next consider a quantum oscillator  initially prepared in thermal equilibrium state with density operator, 
\begin{equation}
\rho_{eq} = \sum_{n =0}^{\infty} p_n^0 \ket{\Psi_n(x,0)} \bra{\Psi_n(x,0)},
\label{s12}
\end{equation} 
where $p_n^0 = \exp(-\beta E_n^0)/Z_0$ is the probability that the oscillator is in  state $\ket{\Psi_n(x,0)}$ and $Z_0$ is the partition function. The initial thermal mean energy at $t = 0$ is accordingly, 
\begin{equation}
\la H(0) \ra = m\omega_0^2 \la x^2(0)\ra = \frac{\hbar \omega_0}{2} \coth\left(\frac{\beta \hbar \omega_0}{2}\right).
\label{s13}
\end{equation}
The  expectation value of the local counterdiabatic Hamiltonian $H_\text{LCD}(t)$ at time $t$ follows from Eqs. (11)-(13) as
 \begin{eqnarray}
 \la H_\text{LCD}(t) \ra &=& \sum_{n=0}^{\infty} p_n^0 \bra{\Psi_n(x,t)} H_\text{LCD} \ket{\Psi_n(x,t)}\nonumber\\ &=& \frac{\la H(0)\ra}{b^2} + \frac{(\dot{b}^2 - b\, \ddot{b})}{2 \omega_0^2} \la H(0)\ra,
\label{s14}
 \end{eqnarray}
where we have used the fact that $\la \{x,p\} (0)\ra = 0$ for  thermal equilibrium state.
Since the squared frequency (\ref{s5}) can be rewritten in the adiabatic limit as,
\begin{equation}
\Omega_t^2 = \omega_t^2 - \frac{\ddot{b}_{ad}}{b_{ad}},
\label{s15}
\end{equation}
we obtain, using Eqs.~(\ref{s5}), (\ref{s8}) and (\ref{s15}), the expression,
\begin{equation}
b\, \ddot{b} - \dot{b}^2 = b^2_{ad} \left\{-\frac{\ddot{\omega}_t}{2 \omega_t} +\frac{1}{2}\left(\frac{\dot{\omega}_t}{\omega_t}\right)^2\right\}.
\label{s16}
\end{equation}
Finally, substituting Eq.~(\ref{s16}) into Eq.~(\ref{s14}), the mean energy of the local counterdiabatic  driving is found to be, 
\begin{equation}
\la H_\text{LCD}(t)\ra= \frac{Q^\ast_\text{LCD} (t)}{b_{ad}^2} \la H(0)\ra,
\label{s17}
\end{equation}
where the adiabaticity parameter $Q^\ast_\text{LCD} (t)$ is given by,
\begin{equation}
 Q^\ast_\text{LCD} (t) = 1 - \frac{\dot{\omega}_t^2}{4 \omega_t^4} + \frac{\ddot{\omega}_t}{4\omega_t^3}.
 \label{s18}
\end{equation}
Note that  Eq.~(\ref{s18}) corrects Eq.~(51) in Ref.~\cite{bea16}.

\begin{figure}[!]
\begin{center}
\includegraphics[width=.95\linewidth]{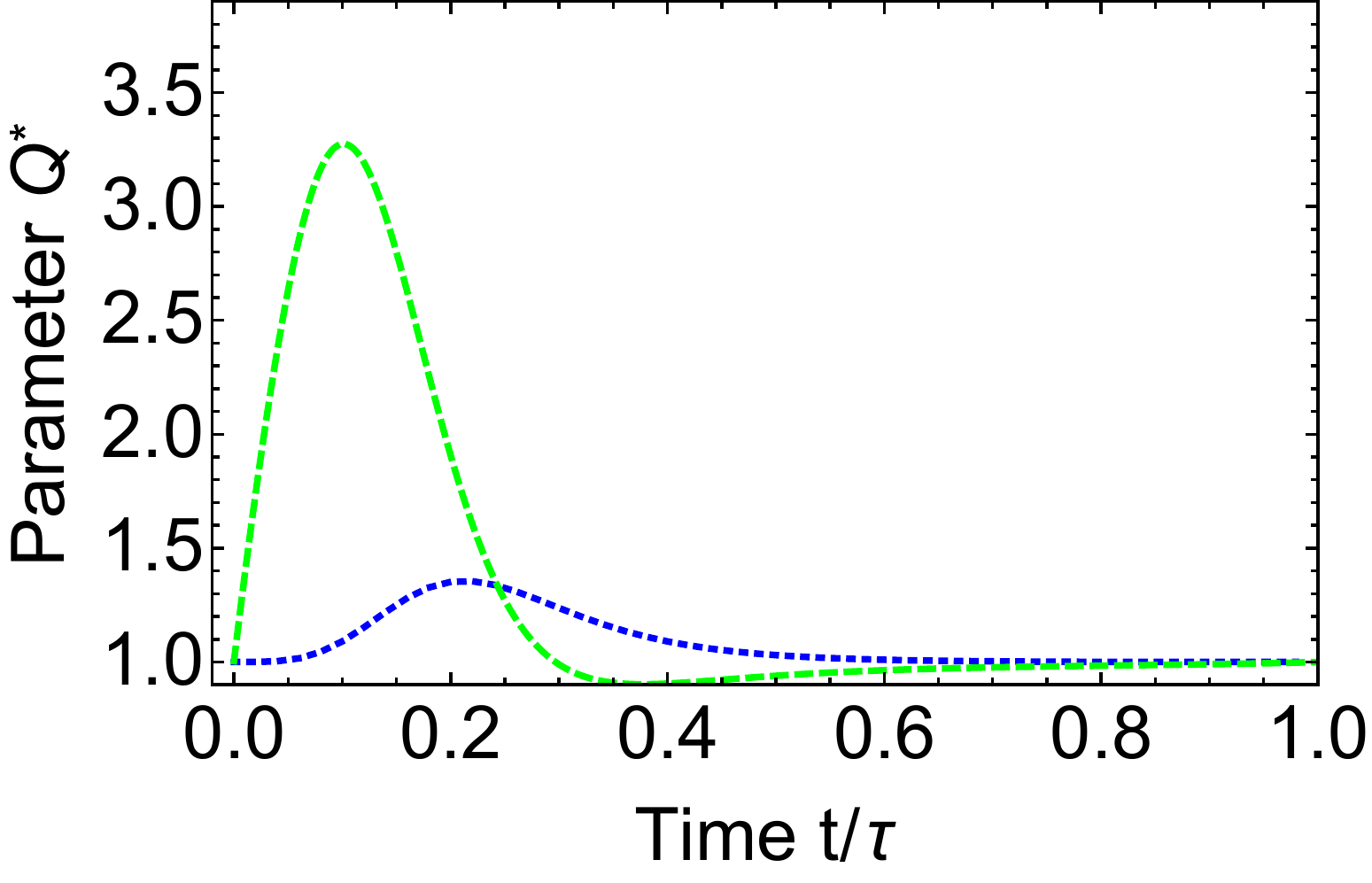}
\end{center}
\caption{Adiabaticity parameter $Q^*_\text{LCD}(t)$, Eq.~(18), (green dashed) and Eq.~(51) of Ref.~\cite{bea16}, $\bar Q^*_\text{LCD}(t) = 1 + \dot \omega_t^2/(8 \omega_t^4$), (blue dotted)  as a function of $t/\tau$ for   $\omega_0/ \omega_1 = 0.15$. }
\label{figS1}
\end{figure}

\section{Superadiabatic energy}
The expectation value of the superadiabatic potential $H_\text{SA}$ may be evaluated from Eqs.~(4) and (17). We have, 
\begin{eqnarray}
 H_\mathrm{SA}(t) &=& H_\mathrm{LCD}(t) - H_\mathrm{0}(t) \nonumber \\ 
 &=& \frac{m}{2}\left(- \frac{3 \dot{\omega}_t^2}{4 \omega_t^2} + \frac{\ddot{\omega}_t}{2\omega_t}\right) x^2.
\end{eqnarray}
As a consequence, we obtain,
\begin{equation}
\la H_\mathrm{SA}(t)\ra = \frac{\omega_t}{\omega_0} \la H(0) \ra\left[- \frac{\dot{\omega}_t^2}{4 \omega_t^4} + \frac{\ddot{\omega}_t}{4 \omega_t^3}\right].
\end{equation}
The properties of the shortcut imply that $\la H_\mathrm{SA}(0,\tau)\ra = 0$ at the beginning and at the end of the protocol.

\section{Bures length}
The Bures length between initial and final density operators of the system is $\mathcal{L}(\rho_\tau,\rho_0) = \arccos \sqrt{F(\rho_\tau,\rho_0)}$, where the $F(\rho_\tau,\rho_0)$ is the fidelity between the two states \cite{scu98}. For the considered driven harmonic oscillator, initial and final states are Gaussian and the fidelity is explicitly given by \cite{def13b}:
\begin{widetext}
\begin{equation}
\mathcal{F}(\rho_\tau,\rho_0) =\frac{2}{\sqrt{ct^2(\beta \epsilon_0/2) + ct^2(\beta \epsilon_1/2) + 2 Q^\ast ct(\beta \epsilon_0/2) ct(\beta \epsilon_1/2) + c^2(\beta \epsilon_0/2)c^2(\beta \epsilon_1/2)} - c(\beta \epsilon_0/2)c(\beta \epsilon_1/2)}.
\end{equation}
\end{widetext}
where $\epsilon_i =  \hbar \omega_i$, $ct(x) = \coth(x)$ and $c(x) = \mathrm{csch}(x)$.
}


\begin{thebibliography}{99}
\bibitem{com} In cases where the system Hamiltonian $H_0$ limits the evolution speed, the superadiabatic approach will only have a restricted applicability.
\end{thebibliography}
\end{document}